\documentclass[aps,prl,twocolumn,superscriptaddress,showpacs]{revtex4}

\usepackage{graphicx}

\begin{document}

\title{Magnetic-Field Dependence of the YbRh$_2$Si$_2$ Fermi Surface}

\author{P.~M.~C. Rourke}
\email{p.rourke@utoronto.ca}
\affiliation{Department of Physics, University of Toronto, Toronto, ON M5S 1A7, Canada}

\author{A. McCollam}
\affiliation{Department of Physics, University of Toronto, Toronto, ON M5S 1A7, Canada}

\author{G. Lapertot}
\affiliation{DRFMC, SPSMS, CEA Grenoble, 17 rue des Martyrs, 3805-4 Grenoble cedex 9, France}

\author{G. Knebel}
\affiliation{DRFMC, SPSMS, CEA Grenoble, 17 rue des Martyrs, 3805-4 Grenoble cedex 9, France}

\author{J. Flouquet}
\affiliation{DRFMC, SPSMS, CEA Grenoble, 17 rue des Martyrs, 3805-4 Grenoble cedex 9, France}

\author{S.~R. Julian}
\affiliation{Department of Physics, University of Toronto, Toronto, ON M5S 1A7, Canada}

\date{\today}

\begin{abstract}
Magnetic-field-induced changes of the Fermi surface play a central role in theories of the exotic quantum criticality of YbRh$_2$Si$_2$. We have carried out de Haas--van Alphen measurements in the magnetic-field range 8 T~$\leq H \leq$~16~T, and directly observe field dependence of the extremal Fermi surface areas. Our data support the theory that a low-field ``large'' Fermi surface, including the Yb 4$f$ quasihole, is increasingly spin split until a majority-spin branch undergoes a Lifshitz transition and disappears at $H_0 \approx 10$~T, without requiring 4$f$ localization at $H_0$.
\end{abstract}


\maketitle

The heavy fermion compound YbRh$_2$Si$_2$ has been the focus of much recent interest because it can be tuned through a field-induced quantum critical point (QCP) which is both easily accessible to experiment and appears to exhibit a new class of ``local'' quantum criticality~\cite{gegnaturephysreview}. It has been argued that this scenario involves dramatic Fermi surface (FS) changes as a function of temperature and magnetic field~\cite{gegnaturephysreview,gegnjp,tokiwaprl}. In this Letter, we have used de Haas--van Alphen (dHvA) measurements to probe these proposed changes.

The main focus of attention in YbRh$_2$Si$_2$ has so far been at low magnetic fields, where the magnitude of the Hall coefficient shows a series of crossovers, extrapolating to a discontinuous jump at the $H_c || c \approx 0.6$~T quantum critical point. This has been interpreted as signalling a sudden Fermi surface reconstruction from a small to a large FS~\cite{paschennature}; such an interpretation is controversial, however, due to the sensitivity of the YbRh$_{2}$Si$_{2}$ Hall effect to small changes in $f$-occupancy~\cite{norman,friedemannphysicab} and/or changes in quasiparticle scattering as antiferromagnetic fluctuations give way to ferromagnetic fluctuations for $H > H_{c}$~\cite{gegfmfluct}. 

At higher fields, specifically at $H_0 \bot c \approx 10$~T, crossovers of dc-magnetization, $T^{2}$ coefficient of resistivity, specific heat, and linear magnetostriction coefficient have been interpreted as evidence for a second Fermi surface reconstruction, back to the small FS, as the Zeeman energy becomes comparable to the Kondo energy scale. This has been suggested to be a continuous change~\cite{gegnjp} or a transition~\cite{tokiwaprl} of the total Fermi volume. 

The primary goal of our work was to look near $H_0$ for this second proposed Fermi surface change. Because dHvA is limited to high fields, we could not directly access the low-field QCP, but the Fermi surface behavior near $H_0$ sheds light on the electronic structure between $H_c$ and $H_0$, and therefore provides experimental constraints for theories of YbRh$_2$Si$_2$ quantum criticality.

\begin{figure}
 \includegraphics[width=3.4in]{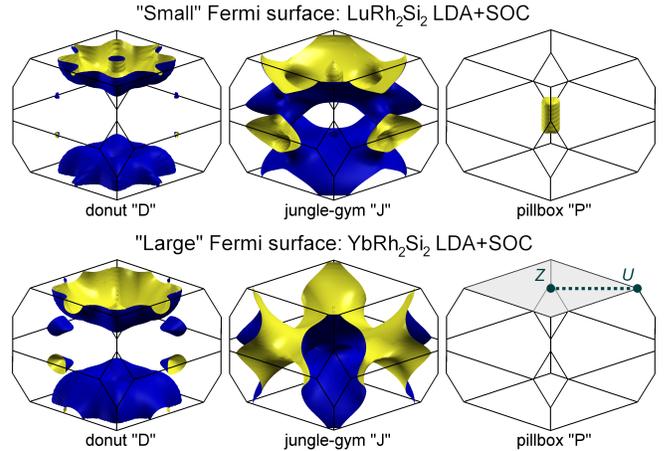}
 \caption{\label{fig:fssheets} (color online) ``Small'' and ``large'' Fermi surfaces calculated with the LDA + spin-orbit coupling method. Following convention from Ref.~\cite{wigger}, the sheets are labeled donut ``\textit{D},'' jungle-gym ``\textit{J},'' and pillbox ``\textit{P}.'' From an electron point of view, the dark (blue online) side of each sheet is the occupied side, and the light (yellow online) side is the unoccupied side, such that \textit{D} is a hole surface and \textit{P} is an electron surface. Note that \textit{D} is a torus in the small Fermi surface case only.}
\end{figure}

The central issue with regard to the proposed Fermi surface reconstructions discussed above is whether or not the Yb 4$f$ quasihole is included in the Fermi volume. In the ``small'' Fermi surface case (also known as ``4$f$ localized'' or ``Yb$^{3+}$''), it does not contribute, whereas in the ``large'' Fermi surface case (also known as ``4$f$ itinerant''), it does. To illustrate these two cases, we have performed all-itinerant LDA + spin-orbit coupling calculations~\cite{lt25proc} for LuRh$_2$Si$_2$ (small FS) and YbRh$_2$Si$_2$ (large FS) using the \textsc{wien2k} density functional theory code~\cite{wien2k}. The large-scale features of our calculated Fermi surfaces, visualized in Fig.~\ref{fig:fssheets} via the \textsc{xcrysden} program~\cite{xcrysden}, are in good agreement with those published previously~\cite{wigger,jeong,knebel,norman}.

\begin{figure}
 \includegraphics[width=3.4in]{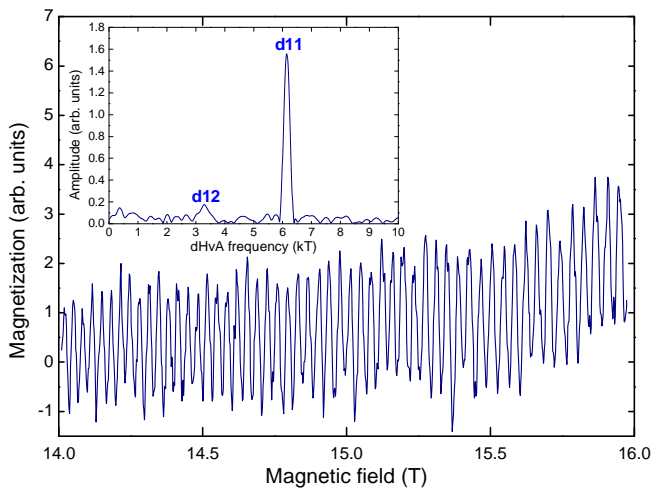}
 \caption{\label{fig:qmosc} (color online) YbRh$_2$Si$_2$ quantum oscillations in the 14--16~T range, with field aligned $\sim$$7^{\circ}$ from (110) toward (001). A Fast Fourier Transform (FFT) in $1/H$ of the displayed data is inset.}
\end{figure}

In order to probe the Fermi surface of YbRh$_2$Si$_2$ as the field is increased through $H_{0}$, we have conducted ambient-pressure dHvA measurements from 8--16~T and 30--600~mK, using the standard field modulation method. The high quality single-crystal samples studied were $\sim$$0.2 \times 1 \times 2$~mm platelets grown from indium flux~\cite{knebel}, with a residual resistivity ratio of about 100. The magnetic field was rotated across a range of $45^{\circ}$ from $a$ = (100) to (110) and $60^{\circ}$ from (110) toward $c$ = (001). Figure~\ref{fig:qmosc} shows a sample trace of the as-measured dHvA oscillations and an associated Fast Fourier Transform (FFT) in the 14--16~T field range.

\begin{figure}
 \includegraphics[width=3.4in]{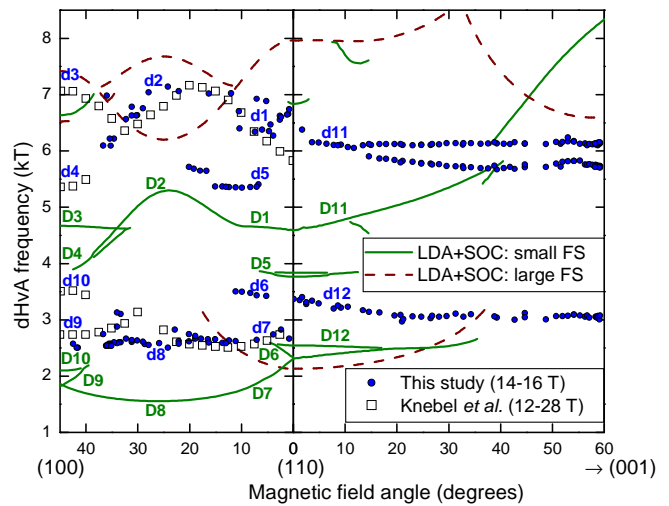}
 \caption{\label{fig:freqvsang} (color online) dHvA frequency vs magnetic-field angle. Closed (blue online) circles depict the current experimental data (14--16~T field range), open squares show previously-published data (12--28~T field range)~\cite{knebel}, and lines are from the calculated small (solid, green online) and large (dashed, red online) Fermi surfaces shown in Fig.~\ref{fig:fssheets}. Calculated orbits from the small FS case are labeled with uppercase (green online) letters and numbers, whereas corresponding experimental orbits are labeled with lowercase (blue online) letters and numbers.}
\end{figure}

Figure~\ref{fig:freqvsang} shows the field-angle dependence of our measured quantum oscillation frequencies in the 14--16~T field range, below 50~mK. For comparison, previously published data from torque measurements on samples grown in the same facility~\cite{knebel}, as well as predicted dHvA frequencies from the calculated small and large Fermi surfaces of Fig.~\ref{fig:fssheets}, are included. Predicted dHvA frequencies and band masses were extracted from each calculated Fermi surface sheet using an algorithm described elsewhere~\cite{skeaf}. With the exception of the extra frequency near 6~kT beyond $\sim$$15^{\circ}$ in the (110)-(001) plane, and persistence of observed frequencies over a larger angular range in this plane, there is reasonable qualitative agreement between the small Fermi surface calculation and experiment. Quantitative agreement is generally not obtained for heavy fermion compounds since the flatness of the bands makes the FS size depend sensitively on small shifts of the Fermi energy. The qualitative agreement between the large Fermi surface calculation and experiment is significantly poorer, since the calculation misses numerous branches in the (100)-(110) plane.

To highlight the correspondence between experiment and theory, matching pairs of orbits have been labeled in Fig.~\ref{fig:freqvsang} with the same letter and number combination---uppercase letters for calculated orbits from the small FS case and lowercase letters for experimental data. Predicted major orbits on the small \textit{J} surface are at higher or lower frequencies and are not shown in Fig.~\ref{fig:freqvsang}, as we were unable to observe them: all of our experimentally detected frequencies correspond to hole orbits on the \textit{D} sheet. Calculated band masses, measured cyclotron effective masses, and corresponding mass enhancements for measured orbits and their theoretical counterparts are listed in Table~\ref{table:mstar}; the average mass enhancement is $10(1)$. Within the 0.1~kT FFT resolution of our experiment, no spin splitting of the dHvA frequencies is observed, and the temperature dependence of all measured oscillations is well described by the standard Lifshitz-Kosevich relation.

 \begin{table}
 \caption{\label{table:mstar} Calculated band masses ($m_b$) from the small Fermi surface in Fig.~\ref{fig:fssheets}, measured effective masses ($m^*$) from the dHvA experiment (14--16~T field range), and resulting mass enhancements ($m^{*}/m_{b}$), for calculated and measured orbits in Fig.~\ref{fig:freqvsang}.}
 \begin{ruledtabular}
 \begin{tabular}{cccccc}
 LDA + SOC orbit & Expt. orbit & $m_b$ ($m_e$) & $m^*$ ($m_e$) & $m^{*}/m_{b}$  \\
 \hline
 \textit{D}1 & \textit{d}1 & 1.00(6) & 10.1(5) & 10.1(8) \\
 \textit{D}2 & \textit{d}2 & 0.82(2) & 8.6(5) & 10.5(7)  \\
 \textit{D}5 & \textit{d}5 & 0.71(3) & 8.6(5) & 12.1(9)  \\
 \textit{D}6 & \textit{d}6 & 0.62(5) & 6.8(3) & 11(1)  \\
 \textit{D}7 & \textit{d}7 & 0.72(1) & 10(1) & 14(1)  \\
 \textit{D}8 & \textit{d}8 & 0.554(4) & 5.00(5) & 9.0(1)  \\
 \textit{D}11 & \textit{d}11 & 1.69(9) & 9.02(9) & 5.3(3)  \\
 \end{tabular}
 \end{ruledtabular}
 \end{table}

\begin{figure}
 \includegraphics[width=3.4in]{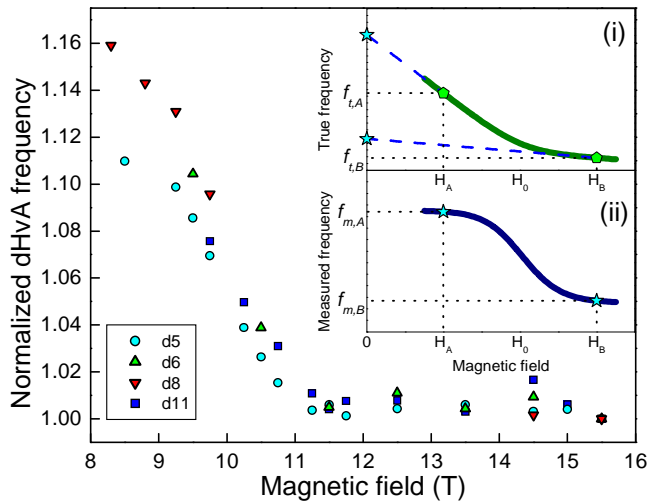}
 \caption{\label{fig:lowfield} (color online) Measured dHvA frequencies as a function of magnetic-field strength, normalized to each orbit's frequency at 15.5~T. Orbit labels correspond to those in Fig.~\ref{fig:freqvsang}; the high-field frequencies for each orbit are: $F_{d5} = 5.4$~kT, $F_{d6} = 3.5$~kT, $F_{d8} = 2.6$~kT, and $F_{d11} = 6.1$~kT. The inset schematically shows the correspondence between a true Fermi volume change (i) and that observed via back-projection in a dHvA experiment (ii). The frequencies we follow lowest in field show the S-shape of inset (ii).}
\end{figure}

As our aim was to investigate possible Fermi surface changes across $H_0$, it is important that we were able to follow the strongest oscillations (\textit{d}5, \textit{d}6, \textit{d}8 and \textit{d}11 from Fig.~\ref{fig:freqvsang}) from 16~T (well above $H_0$) to 8~T (well below $H_0$). Figure~\ref{fig:lowfield} shows the measured frequency of these oscillations as a function of field strength, normalized to the value at 15.5~T. Remarkably, all of the four \textit{D}-surface orbits that we can follow below $H_0$ exhibit the same behavior: roughly constant above about 11~T, rising sharply as the field is reduced through $H_0$, and perhaps tending to saturate below {$\sim$9~T}. 

Interpretation of the field dependence of a measured dHvA frequency, $f_m$, requires some care, because $f_m$ is not directly related to the ``true'' frequency, $f_t$, but rather it is the back-projection to $H = 0$ of the line tangent to $f_t$, as illustrated in the inset of Fig.~\ref{fig:lowfield}~\cite{backproj}. For example, at a particular field $H_A$, $f_{m,A} = f_{t,A} - H_A (\partial f_{t}/\partial H)_A$. Observed field dependence of $f_m$ usually reflects changes of $\partial f_{t}/\partial H$ rather than of $f_t$ itself. In YbRh$_2$Si$_2$, we know that the magnetization roughly saturates above $H_0$~\cite{tokiwaprl}, so the slope $\left|\partial f_{t}/\partial H\right|$ should be lower for $H > H_0$. Thus, the behavior must be roughly as shown in the inset of Fig.~\ref{fig:lowfield}, with a gradual change in $f_{m}$ corresponding to a gradual change in $f_{t}$ and therefore the Fermi surface. We conclude that we are observing a continuous shrinking of the measured \textit{D}-surface at a rate that slows above $H_0$.

It is tempting to infer that the shrinking \textit{D} surface means that the Yb 4$f$ quasihole is gradually disappearing from the Fermi volume, and that $H_0$ represents the completion of this process~\cite{gegnjp}. However, the traditional Luttinger's theorem-based view is that at $T = 0$~K {\em continuous} changes in total Fermi volume should not occur. If the Yb 4$f$ quasihole were to localize across $H_0$, a discontinuous jump in dHvA frequencies would be observed, as in CeRhIn$_5$ under pressure~\cite{shishidodiscontfs}. Since the observed frequencies change continuously across $H_0$, 4$f$ localization must not occur at this field.

An alternative scenario for the behavior at $H_0$ has been developed by Kusminskiy \textit{et al.}~\cite{kusminskiytheory}. They consider the spin splitting of the Fermi surface at high fields, and propose that $H_0$ represents a Lifshitz transition, where a heavy, majority-spin branch of the large Fermi surface disappears, leaving only a moderate-effective-mass minority-spin branch of the large Fermi surface. The 4$f$ localization transition is predicted to occur at much higher fields. This phenomenon occurs in both static~\cite{kusminskiytheory} and dynamic~\cite{beachdmft} mean field theory treatments.

\begin{figure}
 \includegraphics[width=3.4in]{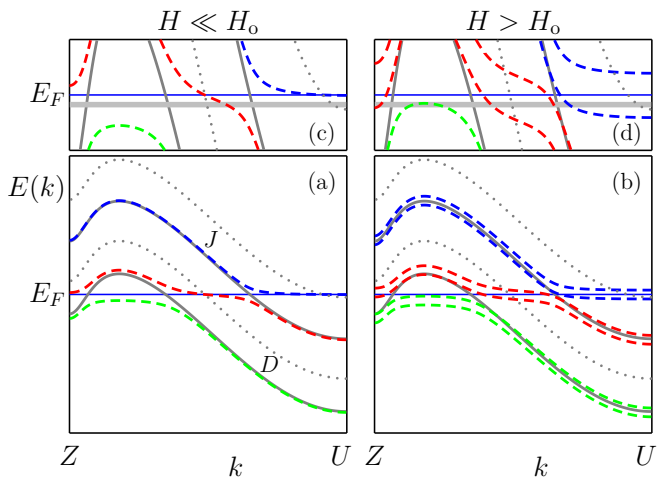}
 \caption{\label{fig:lifshitztransition} (color online) A schematic representation of the YbRh$_2$Si$_2$ bands near the Fermi energy, $E_F$, following Kusminskiy \textit{et al.}~\cite{kusminskiytheory}. Bands associated with the ``\textit{D}'' and ``\textit{J}'' Fermi surface sheets of Fig.~\ref{fig:fssheets} are labeled. Panel (a) shows the band structure without spin splitting: solid gray lines for the small Fermi surface case, and dashed (blue, red and green online) lines for the many-body quasiparticle bands whose Fermi surface coincides with the ``large'' LDA FS (dotted lines). The virtual $f$-hole is shown by a gray horizontal line in panel (c). Panel (b) adds spin splitting to the quasiparticle bands, causing one minority-spin surface to resemble the unsplit small \textit{D} surface, and the largest majority-spin band to no longer cross $E_F$. Panels (c) and (d) show magnified views of panels (a) and (b), respectively, near $E_F$.}
\end{figure}

The Kusminskiy scenario was developed for a toy, one-band model, but we show in Fig.~\ref{fig:lifshitztransition} how it might work for the actual band structure of YbRh$_2$Si$_2$, using a simplified sketch of the \textit{D} and \textit{J} energy bands along the $Z$-$U$ direction in the Brillouin zone. The bands giving rise to the small Fermi surface [solid grey lines in Fig.~\ref{fig:lifshitztransition}(a)] hybridize with a virtual $f$-level near the Fermi energy, $E_F$. The resulting many-body quasiparticle bands [dashed lines in Fig.~\ref{fig:lifshitztransition}(a)] have a {\em large} Fermi surface (i.e., the $f$-hole is now included in the Fermi volume), normally assumed to be the same as the large LDA Fermi surface [dotted grey lines in Fig.~\ref{fig:lifshitztransition}(a)]. When the magnetic field polarizes the quasiparticle bands [dashed lines in Fig.~\ref{fig:lifshitztransition}(b)], the majority- and minority-spin branches split: the minority hole quasiparticle \textit{D}-band sinks, such that its Fermi surface resembles the small \textit{D} FS sheet; the majority hole quasiparticle \textit{D}-band rises to resemble the small \textit{J} FS sheet, where it becomes the opposite-spin counterpart of the minority hole quasiparticle \textit{J}-band; finally, the majority hole quasiparticle \textit{J}-band rises, and $H_0$ is the field at which the quasiparticle Fermi surface corresponding to this band has grown to encompass the entire Brillouin zone and disappears. Thus, through this process, it is possible for the minority branch of the large \textit{D} Fermi surface to resemble the small \textit{D} Fermi surface without requiring 4$f$ localization at $H_0$.

The situation is reminiscent of the heavy fermion compound CeRu$_2$Si$_2$, which exhibits a metamagnetic transition (MMT) at applied fields of $H = (7.7 /\cos \theta_{H})$~T, where $\theta_{H}$ is the angle between the field direction and the crystallographic $c$-axis. dHvA measurements below and above the MMT were matched to band structure predictions for large and small Fermi surface cases respectively, implying a sudden localization of the Ce 4$f$ state at the MMT~\cite{aokiprl}. However, as in YbRh$_2$Si$_2$, measurements of other CeRu$_2$Si$_2$ bulk properties rule out the possibility of a first order transition at the MMT, leading to the proposal of a continuous Fermi surface Lifshitz transition~\cite{daouprl,kusminskiytheory}.

The modified Kusminskiy model discussed above is in accord with both Luttinger's theorem and our experimental results. A key point is that we do not resolve spin splitting of the orbits in any of our measurements, suggesting that what we observe are indeed small-\textit{D}-like oscillations that arise from one spin direction only, as in Fig.~\ref{fig:lifshitztransition}(b). If the 4$f$ quasihole had localized at $H_0$, all of the observed orbits would be spin split. In contrast, in the Kusminskiy model, the majority-spin quasiparticle branch has a completely different topology from the minority-spin branch. As the minority-spin quasiparticle branch shrinks to resemble the small \textit{D} LDA Fermi surface, it becomes a topological torus, allowing the appearance of orbits \textit{D}6, 7, 8, 9, 10, and 12, which thread through the center of the torus. The corresponding majority-spin branch is never toroidal, and thus would not have these orbits. Moreover, as this majority surface grows to resemble the small \textit{J} LDA Fermi surface, its topology will radically change, such that orbits \textit{D}1, 2, 3, 4, and 11 also vanish.

We therefore believe that our result, combined with Luttinger's theorem, rules out the 4$f$ localization scenario and supports the approach of Kusminskiy \textit{et al.}~\cite{kusminskiytheory}: the high-field Fermi surface is a spin split version of the large YbRh$_2$Si$_2$ Fermi surface, rather than a direct realization of the small Fermi surface. This, in turn, implies that the Fermi surface includes the 4$f$ degrees of freedom for $H_c < H < H_0$.

\begin{acknowledgments}
The authors acknowledge useful discussions with S.~E. Sebastian. This work is supported by the Natural
Science and Engineering Research Council of Canada and the Canadian Institute for Advanced Research.
\end{acknowledgments}


\end{document}